\documentclass[10pt,conference]{IEEEtran}
\IEEEoverridecommandlockouts

\usepackage{cite}
\usepackage{amsmath,amssymb,amsfonts}
\usepackage{algorithmic}
\usepackage{graphicx}
\usepackage{textcomp}
\usepackage{xcolor}
\usepackage[hidelinks]{hyperref}
\usepackage{tikz}
\usepackage{mathtools}
\usetikzlibrary{arrows.meta,positioning,fit,calc}

\usepackage[normalem]{ulem}

\newcommand{\op}[1]{\text{\ttfamily\footnotesize #1}}

\def\BibTeX{{\rm B\kern-.05em{\sc i\kern-.025em b}\kern-.08em
    T\kern-.1667em\lower.7ex\hbox{E}\kern-.125emX}}
 
\begin{document}

\title{\LARGE A Trace-Based Assurance Framework for Agentic AI Orchestration: Contracts, Testing, and Governance}

\author{\IEEEauthorblockN{Ciprian Paduraru}
\IEEEauthorblockA{\textit{University of Bucharest}\\
Romania \\
ciprian.paduraru@unibuc.ro}
\and
\IEEEauthorblockN{Petru-Liviu Bouruc}
\IEEEauthorblockA{\textit{University of Bucharest}\\
Romania \\
petru-liviu.bouruc@unibuc.ro}
\and
\IEEEauthorblockN{Alin Stefanescu}
\IEEEauthorblockA{\textit{University of Bucharest \&} \\
\textit{Institute for Logic and Data Science}\\
Romania \\
alin.stefanescu@unibuc.ro}
}

\maketitle

\begin{abstract}
In Agentic AI, Large Language Models (LLMs) are increasingly used in the orchestration layer to coordinate multiple agents and to interact with external services, retrieval components, and shared memory. In this setting, failures are not limited to incorrect final outputs. They also arise from long-horizon interaction, stochastic decisions, and external side effects (such as API calls, database writes, and message sends). Common failures include non-termination, role drift, propagation of unsupported claims, and attacks via untrusted context or external channels.

This paper presents an assurance framework for such Agentic AI systems. Executions are instrumented as Message-Action Traces (MAT) with explicit step and trace contracts. Contracts provide machine-checkable verdicts, localize the first violating step, and support deterministic replay. The framework includes stress testing, formulated as a budgeted counterexample search over bounded perturbations. 
It also supports structured fault injection at service, retrieval, and memory boundaries to assess containment under realistic operational faults and degraded conditions.
Finally, governance is treated as a runtime component, enforcing per-agent capability limits and action mediation (\texttt{allow}, \texttt{rewrite}, \texttt{block}) at the language-to-action boundary.

To support comparative evaluations across stochastic seeds, models, and orchestration configurations, the paper defines trace-based metrics for task success, termination reliability, contract compliance, factuality indicators, containment rate, and governance outcome distributions. 
More broadly, the framework is intended as a common abstraction to support testing and evaluation of multi-agent LLM systems, and to facilitate reproducible comparison across orchestration designs and configurations.
\end{abstract}


\section{Introduction}
Modern enterprise workflows using an Agentic AI approach are increasingly mediated via \emph{LLM-orchestrated multi-agent systems}. In these systems, an orchestration layer decomposes user intent into a plan, delegates subtasks to specialized agents, and executes actions through external services such as APIs, databases, and messaging systems. Production-focused reports highlight that evaluation, reliability, and debugging remain persistent bottlenecks once such systems interact with real services and untrusted inputs \cite{pan_measuring_agents_prod,wef_capgemini_agents}.

In this paper, we focus on \emph{multi-agent LLM system}: multiple LLM-driven agents that interact via messages and shared memory and may invoke external services (e.g., APIs, databases, and messaging).
The coordinating component, termed as \emph{the orchestrator} (orchestration layer), performs routing and role selection, manages shared memory/context, and mediates the translation from language-level decisions into concrete tool actions.

\noindent\textit{Motivating scenario.}
Consider a customer support triage assistant deployed in an enterprise helpdesk. It ingests an inbound email (untrusted text), fetches customer history from an internal database, queries shipping status from a logistics API, drafts a reply, and updates internal tickets. In realistic operation, assurance must account for four recurring stressors:
\begin{itemize}
    \item \textit{Untrusted inputs:} emails or retrieved snippets may include indirect prompt injection attempts.
    \item \textit{Integration faults:} external services may time out, return partial responses, or serve stale cached data.
    \item \textit{Policy constraints:} privacy rules for personally identifiable information (PII) and communication consent rules (e.g., opt-out lists).
    \item \textit{Long-horizon coordination:} handoffs between planner, verifier, and action roles may drift, loop, or deadlock across multi-step workflows.
\end{itemize}

In this context, a test must evaluate not only the final answer but also the sequence of tool actions and side effects under stochastic decisions and imperfect services. This shifts the testing target from output-only oracles to trace-level properties (termination, policy compliance, containment), monitored during execution.

\begin{figure*}[t]
\centering
\includegraphics[width=0.8\textwidth]{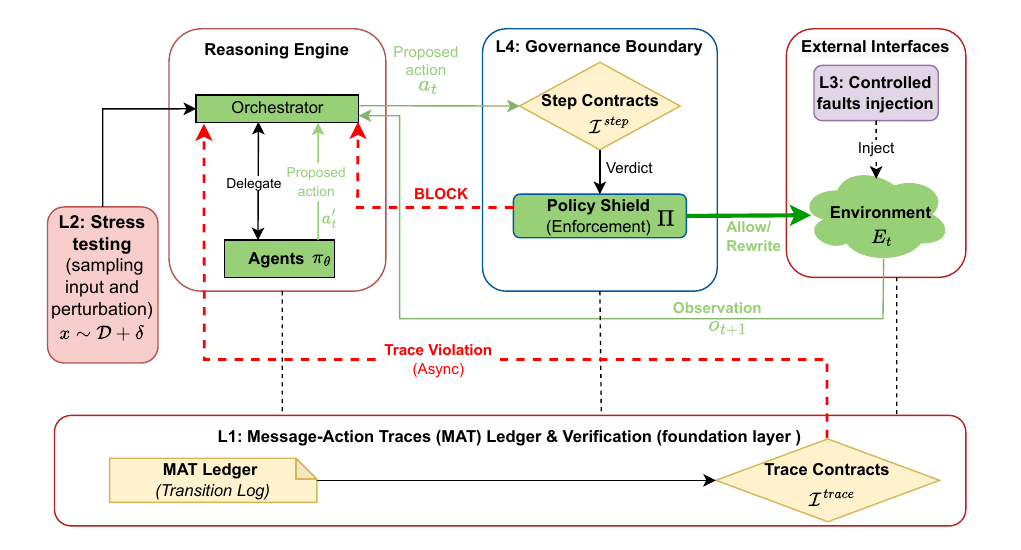}
\caption{\textbf{Pipeline overview of the assurance framework.}
Colors indicate the four layers and their roles.
The \emph{system under test} (SUT, green) is the deployed \emph{multi-agent LLM system}: an orchestrator coordinating an agent pool, together with the runtime governance boundary \textbf{L4} (blue).
The diagram uses a centralized orchestrator for clarity; the same instrumentation and controls apply to decentralized variants (e.g., peer-to-peer agents) by treating the current decision maker as the acting role at step $t$.
The operational environment (green cloud) is depicted only \emph{through} tool, retrieval, and memory interfaces, since the framework evaluates end-to-end integration behavior and integration failures.
\textbf{L2} stress testing (orange) draws task instances $x \sim \mathcal{D}$ and applies bounded perturbations $\delta$ to inputs and context.
During execution, the acting role proposes an action $a_t$; \textbf{L4} mediates the proposed external action using step contracts $\mathcal{I}^{\mathrm{step}}$ and a policy shield $\Pi$, yielding \texttt{allow}, \texttt{rewrite} (to a governed action $\tilde a_t$), or \texttt{block} (red dashed feedback) to prevent unsafe side effects.
\textbf{L3} (red) injects controlled faults at the same external interfaces to exercise realistic integration disturbances.
The resulting observation $o_{t+1}$ (tool output, retrieved evidence, or error) closes the interaction loop.
In parallel, \textbf{L1} (yellow) records Message-Action Trace entries and evaluates trace contracts $\mathcal{I}^{\mathrm{trace}}$ over prefixes, localizing the first violation and emitting a replay record for debugging and regression testing.}

\label{fig:assurance_pipeline}
\end{figure*}

The framework proposed in this work integrates and adapts three directions that are often treated separately:
(i) contract-based runtime monitoring over executions, grounded in runtime verification \cite{rv_survey_sanchez,bauer2011rv};
(ii) robustness testing under realistic perturbations and boundary faults, consistent with chaos engineering and production-oriented agent evaluations \cite{chaos_engineering,pan_measuring_agents_prod};
and (iii) governance controls at the language-to-action boundary, motivated by guidance and analyses on prompt injection, excessive agency, and tool-interface risk \cite{owasp_llm_top10,mcp_security_best_practices,unit42_mcp_attack_vectors}.
Figure~\ref{fig:assurance_pipeline} summarizes how these elements are organized into a single assurance workflow for multi-agent orchestration.

Compared to observability-only tracing or standalone safety checks, the framework treats monitoring, stress testing, and governance as jointly testable through a shared trace-and-contract interface. 
The intent is not to prescribe a specific agent architecture, but to provide a testing-oriented abstraction that can be adopted across implementations to reason about failures, robustness, and governance consistently.

The contributions are summarized below:
\begin{itemize}
    \item \textit{System model and failure taxonomy for multi-agent systems.}
    A lightweight model and taxonomy covering coordination failures (loops and deadlocks), error propagation across agents, role drift in long-horizon workflows, and failures introduced by external services and shared memory.
    
    \item \textit{Message-Action Traces (MAT) as a contract-carrying execution record.}
    A trace representation that records each run as a sequence of typed steps (user/agent messages), tool calls, memory reads and writes, delegation, termination), augmented with provenance and contract verdicts.
    The resulting record supports replay and localization of the first violating step, and serves as the common trace representation used across the pipeline in Fig.~\ref{fig:assurance_pipeline}.

    \item \textit{Stress testing as counterexample search under a budget.}
    A testing formulation that searches for small, realistic perturbation schedules and boundary-fault operators that trigger contract violations under an explicit cost budget. In this context, a \emph{contract} \cite{HECKEL2005145} is a machine-checkable predicate over a step-level MAT record or over a trace (or prefix) that encodes an expected property (e.g., authorization, verification-before-action, progress, containment).
    The outcome is a replayable counterexample schedule together with the first violated contract and a localized trace region. This supports debugging, regression testing, and coverage-guided suite construction over contract IDs and interface/action signatures.

    \item \textit{Governance mechanisms that constrain delegated authority and make outcomes measurable.}
    A set of runtime controls including capability restriction (least privilege via per-agent capability sets) and action-time policy mediation (allow, rewrite, block), with risk-aware routing and escalation to a verifier or human approval step when required by policy.
    Trace-derived metrics are defined to characterize governance behavior, including success and contract-failure rates, containment rate, and distributions over mediator outcomes.
\end{itemize}

The paper focuses on the methodology, definitions, and metrics required to instantiate an executable evaluation protocol; a full empirical study is ongoing but is not reported here.

\section{Related Work}
Research on agentic AI, in particular, multi-agent LLM systems, has progressed quickly in building orchestration frameworks and evaluation tools. Less progress has been made on methods that can \emph{specify, monitor, and enforce} properties of long-running multi-agent executions, especially when external services, retrieval, and shared memory are involved.

\subsection{Frameworks and observability for agentic systems}
AutoGen \cite{autogen} supports multi-agent collaboration through structured conversation patterns. LangGraph \cite{langgraph} offers stateful workflows with cycles, retries, and human checkpoints. For debugging and evaluation, LangSmith \cite{langsmith} records intermediate steps and provides scoring and analysis tools. For retrieval-augmented generation, RAGAS \cite{ragas} provides metrics such as faithfulness and context relevance. 
These systems make development and diagnosis easier, but they typically do not enforce properties such as termination, invariant preservation, or role stability under stochastic execution.

\subsection{Benchmarks for capability, security, and misuse}
AgentBench \cite{agentbench} and GAIA \cite{gaia} evaluate task competence in interactive and assistant-style settings. They are useful for measuring what agents can do, but they are not designed to systematically test robustness under small perturbations or to expose multi-agent failure patterns such as delegation loops and error cascades.
Security-focused benchmarks study indirect prompt injection in agent settings. BIPIA \cite{bipia} targets indirect prompt injection for LLM applications. AgentDojo \cite{agentdojo} provides realistic tasks with attack and defense variants. WASP \cite{wasp} studies web agents under hijacking-style objectives. InjecAgent \cite{injecagent_acl} evaluates prompt injection threats and defenses for agents integrated with external services. Misuse-oriented benchmarks extend evaluation beyond benign reliability. AgentHarm \cite{agentharm} measures harmful behavior in multi-step tasks. HarmBench \cite{harmbench} evaluates refusal and robustness to adversarial prompting.

The study in \cite{llmfails} provides an empirical characterization of breakdowns in multi-agent LLM workflows, emphasizing recurrent patterns such as coordination and non-termination behaviors, verification and oversight gaps, role-level drift over long horizons, and error propagation across agents.
Our failure taxonomy (cf.\ F1-F5 from Sec.~\ref{sec:taxonomy}) is aligned with these observations but is structured to support monitored, trace-level evaluation: F1 captures coordination collapse and non-termination; F2 formalizes unsupported-claim propagation as a trace-conditioned phenomenon; and F3 captures role drift as boundary violations against role-specific constraints. In addition, our taxonomy makes two dimensions explicit that are not central in \cite{llmfails}: (a) interface-driven compromise and poisoning via external services, retrieval, or shared memory (F4), motivated by prompt-injection and excessive-agency analyses; and (b) misuse-oriented harmful task execution (F5), motivated by safety and misuse evaluations. This mapping positions \cite{llmfails} as an empirical basis for what fails, while our taxonomy specifies failure conditions in a form suitable for contracts, stress testing, and governance measurement.

\subsection{Runtime constraints, monitoring, and interface risk}
Contract-based testing has been studied in the context of web services \cite{HECKEL2005145}, where contracts specify admissible input-output behavior and interaction protocols. We generalize this notion to agentic AI systems by specifying \textit{contracts} a machine-checkable boolean predicate over an agentic execution, defined on individual steps or finite trace prefixes (e.g., “tool calls with side effects must be preceded by a verification action within $h$ steps”). Contracts allow temporal, cross-agent, and interface-level properties to be specified independently of model internals.

NeMo Guardrails \cite{nemo_guardrails} provides programmable mechanisms to apply policy constraints, often as dialogue and safety controls. Runtime verification offers a formal basis for monitoring executions against specifications \cite{rv_survey_sanchez,bauer2011rv}, motivating contract-based monitoring for complex workflows.
External service interfaces introduce additional risk. The Model Context Protocol (MCP) specification \cite{mcp_spec} and its security best practices \cite{mcp_security_best_practices} discuss trust and safety concerns when connecting models to tools and data. Unit 42 \cite{unit42_mcp_attack_vectors} reports prompt injection-style attack vectors in MCP deployments. These results motivate capability limits and policy mediation at the point where an internal plan can trigger an external side effect.

\noindent\textit{Failure modes and their operationalization.}
Recent analyses and benchmarks collectively highlight recurrent failure patterns in agentic systems, including non-termination and coordination breakdowns, error amplification across steps, role confusion, injection and interface poisoning, and harmful task completion \cite{llmfails,owasp_llm_top10,bipia,agentdojo,wasp,injecagent_acl,agentharm,harmbench,mcp_security_best_practices,unit42_mcp_attack_vectors}. Our taxonomy in Sec.~\ref{sec:taxonomy} aligns these observations with trace-level conditions that are directly monitorable: coordination and termination failures (\textbf{F1}), unsupported-claim propagation (\textbf{F2}), role drift and boundary violations (\textbf{F3}), interface-driven injection (\textbf{F4}), and misuse outcomes (\textbf{F5}). The goal is not to re-categorize prior work, but to provide a compact set of monitored failure classes that can be exercised by stress testing and fault injection and measured under governance controls.

\subsection{Gap}
Most existing work focuses on either building agent systems or observing them after the fact. What is missing is a single assurance approach that combines:
(i) traces enriched with contracts that can be checked at runtime,
(ii) stress testing that searches for small, realistic perturbations that cause contract violations,
(iii) fault injection across external services, retrieval, and shared memory,
and (iv) governance that limits authority and routes high-impact actions through policy and, when needed, human approval.
This paper proposes such a framework to narrow the existing gaps.

\section{System Model and Failure Taxonomy for Agentic Systems}
\label{sec:taxonomy}

We summarize prior reports of agentic failure patterns into five operational classes, each defined as a trace-level condition that can be monitored and stress-tested.

\subsection{System model}
An agentic system can be viewed as a stochastic transition system with an explicit external environment and a mutable shared state, coordinated via an orchestrator. The orchestrator routes between agents, manages shared memory and context, and mediates external tool actions.
Let $\mathcal{A}=\{1,\dots,m\}$ be the set of agents.. At step $t$, the global state $s_t \in \mathcal{S}$, Eq.~(\ref{eq:global-state}), summarizes:
\begin{itemize}
    \item shared memory $M_t$,
    \item per-agent local contexts $C_t^i$ (role prompt, conversation, working notes),
    \item orchestration control state $G_t$ (for example the current node in an execution graph, retry counters, budgets),
    \item external environment state $E_t$ (for example service sessions, database state, network conditions).
\end{itemize}

\begin{equation}
\label{eq:global-state}
s_t \;=\; (M_t,\; C_t^1,\dots,C_t^m,\; G_t,\; E_t).
\end{equation}
In practice, assurance instrumentation stores only a filtered projection of this state (for example IDs, hashes, and redacted parameters) that is sufficient for monitoring, which reduces retention of sensitive prompt and service content.

An action $a_t \in \mathcal{U}$ denotes an externally meaningful step such as emitting a message, calling an external service with parameters, reading or writing memory, delegating to another agent, or terminating a run. Executing $a_t$ yields an observation $o_{t+1}$ (service response, retrieved passages, error code) and a next state sampled from the environment dynamics:
\[
s_{t+1} \sim P(\cdot \mid s_t, a_t), \qquad a_t \sim \pi_{\theta}^{i_t}(\cdot \mid C_t^{i_t}, M_t),
\]
where $i_t$ is the selected agent at step $t$ and $\pi_{\theta}^{i}$ denotes the stochastic policy induced by an LLM under role instructions and prompt context. This view emphasizes that failures arise from non-deterministic, long-horizon interaction (sampling, external service nondeterminism, partial observability), rather than fixed control flow \cite{llmfails,pan_measuring_agents_prod}.

One execution is represented as a finite trace:
\[
\tau \;=\; (s_0,a_0,o_1,s_1,\dots,s_T),
\]
where $T$ is the configured horizon or termination time. Let $\mathrm{Term}(\tau)=1$ if a terminal action occurs within $T$ steps (successful completion, safe abort, or explicit failure return). Failures are violations of properties over steps or over the whole trace. Two cases are distinguished:
\begin{itemize}
    \item local step failures, where a single transition violates a constraint,
    \item workflow failures, where termination, safety, or correctness properties are violated over the trace.
\end{itemize}
This aligns with runtime verification, which instruments executions and monitors traces against specifications \cite{rv_survey_sanchez,bauer2011rv}.

\subsection{Failure taxonomy}
Recent empirical work reports fine-grained multi-agent failure modes that include specification and system design issues, inter-agent misalignment, and breakdowns in verification and termination \cite{llmfails}.
These are compressed into operational classes that can be tested under the trace model above, and extended with security-focused classes emphasized by agent security benchmarks and guidance \cite{agentdojo,wasp,ncsc_prompt_injection_worse}.

\noindent\textit{F1: Coordination collapse and non-termination.}\\
\noindent\textit{Operational definition:} the system fails to make progress and does not terminate within the configured horizon, due to deadlock, oscillation, or circular delegation.
Termination and verification failures are a major category in multi-agent systems \cite{llmfails}. Common subtypes include circular delegation, deadlock on approvals, repeated replanning without execution, and collisions in shared memory. Orchestrators that support cycles and retries increase expressiveness, but also increase the need for termination and invariant checks \cite{langgraph}.
In the framework, collapse is detected using a non-negative potential function 
$\Phi:\mathcal{S}\rightarrow\mathbb{R}_{\ge 0}$ that encodes remaining work using observable orchestration signals possibly weighted (for example $\Phi(s_t)=\#\text{unresolved subtasks}+\#\text{pending approvals}+\#\text{active retries}$). Collapse is detected as a sufficient condition over a sliding window of length $w$ when the potential does not decrease, and the system does not terminate:
\begin{equation}
\label{eq:coord-collapse}
\forall\, t \in [k,k+w]:\;
\Phi(s_{t+1}) \ge \Phi(s_t)
\;\wedge\;
\neg \mathrm{Term}(\tau).
\end{equation}

\noindent\textit{F2: Error amplification and unsupported-claim propagation.}\\
\noindent\textit{Operational definition:} an upstream factual error is accepted as an assumption and subsequently drives downstream actions or final outputs.
A local factual error can become a system-level failure when an upstream output is treated as authoritative by downstream agents or by the orchestrator \cite{llmfails}.
Let the final response be decomposed into atomic claims $\{c_j\}$ (minimal verifiable propositions), extracted by a lightweight claim splitter (rule-based or model-assisted). Each claim is linked to provenance evidence $\{e_j\}$ recorded in the trace (retrieved passages, external service outputs, database row identifiers). Define $\mathrm{support}(c_j,e_j)=1$ when the evidence provides sufficient justification for the claim under a chosen verifier (e.g., entailment checking, exact result matching, or a constrained LLM judge). Otherwise, $\mathrm{support}(c_j,e_j)=0$. The unsupported claim rate is:
\[
H_{\mathrm{rate}}=\frac{|\{j: \mathrm{support}(c_j,e_j)=0\}|}{|\{j: c_j\}|}.
\]
To capture propagation, mark whether an unsupported claim becomes an input assumption for later actions (e.g., for example service calls, approvals, or downstream instructions). Let $\mathrm{use}(c_j,\tau)=1$ if claim $c_j$ is referenced in subsequent steps. Then:
\[
H_{\mathrm{prop}}=\frac{|\{j:\mathrm{support}(c_j,e_j)=0 \wedge \mathrm{use}(c_j,\tau)=1\}|}{|\{j: c_j\}|}.
\]

Propagation is most harmful when high-impact claims are not re-checked against external services or retrieval before driving actions \cite{pan_measuring_agents_prod}.

\noindent\textit{F2: Error amplification and unsupported-claim propagation.}\\
\noindent\textit{Operational definition:} an upstream factual error is accepted as an assumption and subsequently drives downstream actions or final outputs.
A local factual error can become a system-level failure when an upstream output is treated as authoritative by downstream agents or by the orchestrator \cite{llmfails}.

Let the final response be decomposed into atomic claims $\{c_j\}$ (minimal verifiable propositions), extracted by a lightweight claim splitter (rule-based or model-assisted).
Each claim is linked to provenance evidence $\{e_j\}$ recorded in the MAT trace (retrieved passages, external service outputs, database row identifiers).
Define $\mathrm{support}(c_j,e_j)=1$ when the evidence provides sufficient justification for the claim under a chosen verifier (e.g., entailment checking, exact result matching, or a constrained LLM judge), and $\mathrm{support}(c_j,e_j)=0$ otherwise. 
Let $\mathcal{C}=\{1,\dots,J\}$ index the extracted claims.
The unsupported claim rate is:
\[
H_{\mathrm{rate}}
=\frac{1}{J}\,|\{j\in\mathcal{C}:\mathrm{support}(c_j,e_j)=0\}|.
\]

To capture propagation, mark whether an unsupported claim becomes an input assumption for later actions (e.g., service calls, approvals, or downstream instructions).
Let $\mathrm{use}(c_j,\tau)=1$ if claim $c_j$ is referenced in subsequent steps.
Then:
\[
H_{\mathrm{prop}}
=\frac{1}{J}\,|\{j\in\mathcal{C}:\mathrm{support}(c_j,e_j)=0 \wedge \mathrm{use}(c_j,\tau)=1\}|.
\]
Propagation is most harmful when high-impact claims are not re-checked against external services or retrieval before driving actions \cite{pan_measuring_agents_prod}.


\noindent\textit{F3: Role drift and boundary violations.}\\
\noindent\textit{Operational definition:} an agent deviates from its assigned role by taking unauthorized actions or failing role-specific obligations over the execution horizon.
Role confusion arises when an agent deviates from its declared role or when role boundaries collapse \cite{llmfails}.
Let each agent $i$ have a role contract $\mathcal{R}_i=(\mathcal{U}_i,\mathcal{O}_i)$ specifying allowed actions $\mathcal{U}_i\subseteq\mathcal{U}$ and obligations $\mathcal{O}_i$ (e.g., cite sources, only propose external calls, or avoid external side effects). A role violation occurs when $a_t\notin \mathcal{U}_{i_t}$. Obligation violations are captured by explicit contracts (Sec.~\ref{sec:framework}, L1). A compact \emph{per-agent} trace-level drift score is:
\[
D_i(\tau)=\frac{1}{|\mathcal{T}_i|}\sum_{t\in\mathcal{T}_i}\mathbb{I}\{a_t\notin \mathcal{U}_i\},
\]
where $\mathbb{I}\{\cdot\}$ denotes the indicator function and $\mathcal{T}_i=\{t:\, i_t=i\}$. Drift becomes more likely as context accumulates, instructions interfere, or injected content changes effective constraints.

\noindent\textit{F4: Tool/memory injection and interface poisoning.}\\
\noindent\textit{Operational definition:} untrusted text entering via memory or external interfaces alters the agent policy so that it selects an unsafe action.
Shared memory and external interfaces create direct attack surfaces. OWASP \cite{owasp_llm_top10} identifies \textit{Prompt Injection} and highlights \textit{Excessive Agency}.
Benchmarks and analyses show that prompt injection against agents that call external services is practical and defenses remain incomplete \cite{bipia,agentdojo,wasp,injecagent_acl}.
MCP security guidance recommends treating service integration as a system design problem, with explicit limits and checks, rather than relying on prompt-based defenses alone \cite{mcp_security_best_practices}.

Injection is modeled as a bounded adversarial perturbation $\delta$ applied to untrusted content entering the system via memory or external interfaces (e.g., retrieved text, tool outputs, or stored notes).
Let $M'_t=\mathrm{inject}(M_t,\delta)$ denote a poisoned memory state (similarly for poisoned interface descriptions or metadata). The attack succeeds if the induced policy selects an unsafe action $a_t \in \mathcal{U}_{\mathrm{unsafe}}$, where $\mathcal{U}_{\mathrm{unsafe}}$ denotes actions that violate security or safety policy (e.g., exfiltrating sensitive data, unauthorized service use, or irreversible external actions).

Excessive Agency is the condition where the system is granted enough permissions that plausible perturbations or ordinary service failures can lead to unsafe external actions. In the trace model, a simple operational signal is:
\begin{equation}
\label{eq:excessive-agency-prob}
\Pr(a_t \in \mathcal{U}_{\mathrm{unsafe}}) > \epsilon,
\end{equation}
for some small but meaningful $\epsilon$ within the allowed perturbation budget. Standardized integration protocols can increase exposure by making it easier to connect many services and endpoints \cite{mcp_spec,mcp_security_best_practices,unit42_mcp_attack_vectors}.

\noindent\textit{F5: Misuse and harmful task execution.}\\
\noindent\textit{Operational definition:} the system completes a harmful multi-step objective or performs a harmful external action without triggering refusal, containment, or escalation.
Misuse-oriented benchmarks motivate explicit evaluation along this dimension \cite{agentharm,harmbench}. Operationally, a misuse failure is recorded when the run reaches a harmful end state (or performs a harmful external action) while governance and contracts remain unsatisfied or bypassed.

\section{Framework: Trace Contracts, Adversarial Testing, and Governance}
\label{sec:framework}
\noindent

This section presents a trace-based assurance framework for multi-agent LLM systems that interact with external services, retrieval, and shared memory. 
Layers L1--L3 form the surrounding assurance harness used to observe and stress-test the SUT.
The deployed \emph{System Under Test (SUT)} comprises the orchestrator, agent pool, and the runtime governance mediator (L4), since these jointly determine which external actions are executed. Layers L1–L3 constitute the assurance harness that instruments execution, applies perturbations/faults at interfaces, and checks contracts; the external environment is exercised through tool/retrieval/memory interfaces.

Figure~\ref{fig:assurance_pipeline} provides a pipeline view of how the assurance harness exercises and audits executions.
A task instance $x$ is executed under configuration $\kappa$ (roles, topology, contracts, and governance settings) and stochastic seed $z$, while the harness applies a perturbation or fault schedule $\delta$.
The run emits Message-Action Trace records that are checked against step and trace contracts, producing localized violations and a replay record for debugging and regression testing.
Layer definitions correspond to Sec.~\ref{sec:l1}--~\ref{sec:l4}.

\noindent\textit{Layered responsibilities.}
The responsibilities of the four layers are:
(i) \textbf{L1} instruments each run as Message-Action Trace records with provenance and step/trace contract verdicts, enabling monitoring, localization of the first violating step, and replay.
(ii) \textbf{L2} performs stress testing by searching for low-cost perturbation schedules $\delta$ (and counterexamples $\delta^\star$) that trigger contract violations under a budget, as in Eq.~\ref{eq:min-counterexample}.
(iii) \textbf{L3} injects structured boundary faults at external interfaces (services, retrieval, memory) using a fault schedule and checks the containment requirement (Eq.~\ref{eq:containment}).
(iv) \textbf{L4} governs external actions at the language-to-action boundary by enforcing per-agent capability sets $\mathcal{K}_i$ and mediating tool calls via the policy shield $\Pi$ (allow / rewrite / block), as in Eq.~\ref{eq:policy-shield}.


\subsection{Assurance as monitored traces under perturbations}
Assurance for agentic systems can be formulated as \emph{trace-based} verification under environment perturbations and tool faults. Let $x$ denote a task instance and let $\tau(x,\delta,\Pi)$ be the resulting execution trace when the system is exposed to a perturbation schedule $\delta$ and mediated by a governance policy $\Pi$. 
Let $K_s = |\mathcal{I}^{\mathrm{step}}|$ and $K_\tau = |\mathcal{I}^{\mathrm{trace}}|$ denote the number of step and trace contracts, respectively.
The setting assumes:
\begin{itemize}
    \item a finite set of \emph{step contracts} $\mathcal{I}^{\mathrm{step}}=\{I^{\mathrm{step}}_1,\dots,I^{\mathrm{step}}_{K_s}\}$ evaluated on Message-Action Traces records (local invariants);
    \item a finite set of \emph{trace contracts} $\mathcal{I}^{\mathrm{trace}}=\{I^{\mathrm{trace}}_1,\dots,I^{\mathrm{trace}}_{K_\tau}\}$ evaluated on full traces or prefixes (workflow/temporal properties);
    \item a perturbation/operator space $\Delta$ capturing prompt ambiguity, tool latency/failures, retrieval noise, and memory/tool-channel injection;
    \item a governance policy $\Pi$ that mediates tool execution (allow/rewrite/block) and may impose capability constraints.
\end{itemize}

Let $\mathcal{T}$ denote the set of finite MAT traces emitted by the instrumented system under a fixed horizon $T$. 
A run fails if any \emph{trace contract} is violated (i.e., violation of a single hard safety or correctness constraint is sufficient):

\begin{equation}
\label{eq:fail-def}
\mathrm{Fail}(\tau,\mathcal{I}^{\mathrm{trace}}) \;=\;
\mathbb{I}\Big\{\exists k \in \{1,\dots,K_\tau\}:\ I^{\mathrm{trace}}_k(\tau)=0\Big\}.
\end{equation}

Assurance can be posed as \emph{counterexample search} with an explicit perturbation cost: find the lowest-cost $\delta\in\Delta$ that causes at least one trace contract to fail,
\begin{equation}
\label{eq:min-counterexample}
\delta^\star \in \arg\min_{\delta \in \Delta}\ \mathrm{cost}(\delta)
\quad \text{s.t.} \quad
\mathrm{Fail}\!\big(\tau(x,\delta,\Pi),\mathcal{I}^{\mathrm{trace}}\big)=1.
\end{equation}

When $\delta$ is a schedule $\delta_{0:T}$, the total cost is aggregated as
\[
\mathrm{cost}(\delta)=\sum_{t=0}^{T}\mathrm{cost}(\delta_t).
\]

\subsection{L1: MAT as contract-enriched instrumentation} \label{sec:l1}
Standard logs capture events but rarely provide \emph{semantic accountability}: which role made which claim, based on which evidence, under which constraints, and with what verification outcome. Executions are instrumented as \emph{Message-Action Traces}, where each step is a typed record enriched with (i) provenance and (ii) explicit, checkable contracts. This follows runtime verification (monitor executions against specifications), adapted to agentic workflows where language, tools, and memory jointly determine behavior \cite{rv_survey_sanchez,bauer2011rv}.

\noindent\textit{MAT record and step semantics.}
At step $t$, a controller (centralized or distributed) selects an acting agent $i_t$ and produces an action $a_t$ (message, tool call, memory update, delegation, or termination). The environment returns an observation $o_{t+1}$ (tool result, retrieved context, user reply), and the system updates its internal state. The system emits:
\begin{equation}
\label{eq:mat-record}
r_t = \langle
t,\ i_t,\ \mathrm{role}(i_t),\ \hat{s}_t,\ a_t,\ o_{t+1},\ \mathrm{prov}_t,\ \mathcal{I}^{\mathrm{step}}_t,\ \mathrm{verdict}_t
\rangle .
\end{equation}
Here, $\mathcal{I}^{\mathrm{step}}_t$ is the set of step contracts checked at time $t$ (selected based on the action type). The field $\mathrm{verdict}_t$ stores the result of these checks, e.g., whether the step passed and which contract IDs were violated.

\noindent\textit{Provenance for factuality and audit.}
Let $\mathrm{prov}_t$ be the set of provenance links recorded at step $t$. 
Each link has the form $(\texttt{src}, \texttt{rel}, \texttt{dst})$, where
\texttt{src} and \texttt{dst} are trace artifacts (IDs for claim, retrieved passage,
tool call, tool result, or memory entry), and \texttt{rel} is the link type (e.g., \textit{supports}, \textit{returns}, \textit{derived\_from}).

\noindent\textit{Contract interface and selection.}
Let $\mathcal{R}$ denote the space of MAT records of the form in Eq.~(\ref{eq:mat-record}). Step contracts are boolean predicates on the current record, 
$I^{\mathrm{step}}_k:\ \mathcal{R}\rightarrow\{0,1\}$, while trace contracts are evaluated on the trace (or a prefix), $I^{\mathrm{trace}}_k:\ \mathcal{T}\rightarrow\{0,1\}$. 
At each step, a relevant subset $\mathcal{I}^{\mathrm{step}}_t \subseteq \mathcal{I}^{\mathrm{step}}$ is evaluated, determined by the action type (e.g., tool calls trigger policy checks; memory writes trigger sanitization; final responses trigger factuality/PII checks). 
Trace contracts in $\mathcal{I}^{\mathrm{trace}}$ are checked on prefixes $\tau_{0:t}$ and on termination.

At each step, a relevant subset $\mathcal{I}^{\mathrm{step}}_t \subseteq \mathcal{I}^{\mathrm{step}}$ is evaluated, determined by the action type (e.g., tool calls trigger policy checks; memory writes trigger sanitization; final responses trigger factuality and PII checks, such as detection of personal identifiers against policy-specific allowlists (permitted identifiers) or redaction rules (masking or removal when disclosure is not allowed).
Trace contracts in $\mathcal{I}^{\mathrm{trace}}$ are checked on prefixes $\tau_{0:t}$ and on termination.

\noindent\textit{Verdicts and localization.}
Monitoring outcomes are recorded as:
\begin{equation}
\label{eq:verdict}
\mathrm{verdict}_t=
\langle \mathrm{pass}_t,\ \mathrm{violations}_t,\ \mathrm{severity}_t,\ \mathrm{mitigation}_t\rangle,
\end{equation}
where $\mathrm{violations}_t$ is the set of violated contract IDs, $\mathrm{severity}_t$ is a discrete level (e.g., soft vs.\ hard), and $\mathrm{mitigation}_t$ records the response (e.g., retry, replan, sandbox, escalate, block). This localizes failures to specific steps and agents and supports replay.

\noindent\textit{Example base contract templates.}
Contracts can be drawn from a small library of templates instantiated with system-specific parameters (e.g., tool sets, allowlists, and window sizes):
\begin{itemize}
    \item \emph{Verify before acting:} any side-effecting external call must be preceded by a verifier step within the last $h$ steps.
    \item \emph{Principle of least privilege:} an external service $\mathsf{T}$ may be invoked only if it is permitted by $\mathcal{K}_{i_t}$ and the call parameters satisfy predefined allowlists.
    \item \emph{Progress:} $\Phi$ must decrease at least once in any window of length $w$, unless the run terminates (Eq.~\ref{eq:coord-collapse}).
\end{itemize}
These templates aim to cover correctness (progress and termination), safety (verification gates), and security (capability limits).

\subsection{L2: Adversarial stress testing as constrained environment search} \label{sec:l2}
Capability benchmarks primarily measure average case task performance and contract compliance, rather than behavior under perturbations and faults. Assurance also needs \emph{counterexample discovery}: finding small, plausible perturbations that cause contract violations under an explicit budget. Stress testing is therefore posed as a constrained search over the main input and interaction channels: prompt and context, external services, retrieval, and memory. In this paper, \emph{adversarial}, refers to worst-case perturbation selection within a bounded, plausible operator set $\Delta$ under a cost budget $B$, not necessarily a malicious human attacker.

\noindent\textit{Perturbation schedules.} A perturbation $\delta$ can be applied once (e.g., rewriting the user request) or applied over time as a schedule. Let $\delta_{0:T}=(\delta_0,\ldots,\delta_T)$ with $\delta_t\in\Delta$. The resulting execution trace is
\[
\tau \;=\; \mathrm{Exec}(x,\kappa,z,\delta_{0:T}),
\]
where $\kappa$ is the system configuration (roles, topology, services, governance) and $z$ is the stochastic seed. We write $\delta_{0:T}=(\delta_0,\ldots,\delta_T)$ for a perturbation schedule; when the per-step form is not needed, we denote the schedule simply by $\delta$.

\noindent\textit{Violation signal and plausibility cost.}
At step $t$, only the step contracts relevant to the current action are evaluated. Let
$\mathcal{I}^{\mathrm{step}}_t \subseteq \mathcal{I}^{\mathrm{step}}$
denote this selected subset (e.g., contracts for external service calls, memory writes, or the final response).
Given the MAT record $r_t$, define a weighted violation score:
\[
\mathrm{Vio}(r_t,\mathcal{I}^{\mathrm{step}}_t)
\;=\;
\sum_{I^{\mathrm{step}}_k\in \mathcal{I}^{\mathrm{step}}_t}
\alpha_k \cdot \mathbb{I}\!\left\{I^{\mathrm{step}}_k(r_t)=0\right\},
\]
where $\alpha_k \ge 0$ encodes contract severity.

Each perturbation operator $\delta_t$ is assigned an explicit cost that reflects how intrusive it is:
\begin{equation}
\label{eq:op-cost}
\mathrm{cost}(\delta_t)=
c_{\mathrm{tok}}\,|\Delta\mathrm{tokens}|
+c_{\mathrm{hook}}\,n_{\mathrm{hooks}}
+c_{\mathrm{mag}}\,\eta .
\end{equation}
Here $|\Delta\mathrm{tokens}|$ counts token changes in prompts or other text inputs; $n_{\mathrm{hooks}}$ counts activated boundary fault hooks (e.g., forcing a timeout or delay on a specific external call); and $\eta$ is a magnitude parameter for response perturbation (e.g., removing or changing fields in a structured payload). A fixed budget $B$ bounds the total cost so perturbations remain small, realistic, and comparable across runs.

\begin{figure*}[t]
\centering
\includegraphics[width=0.99\textwidth]{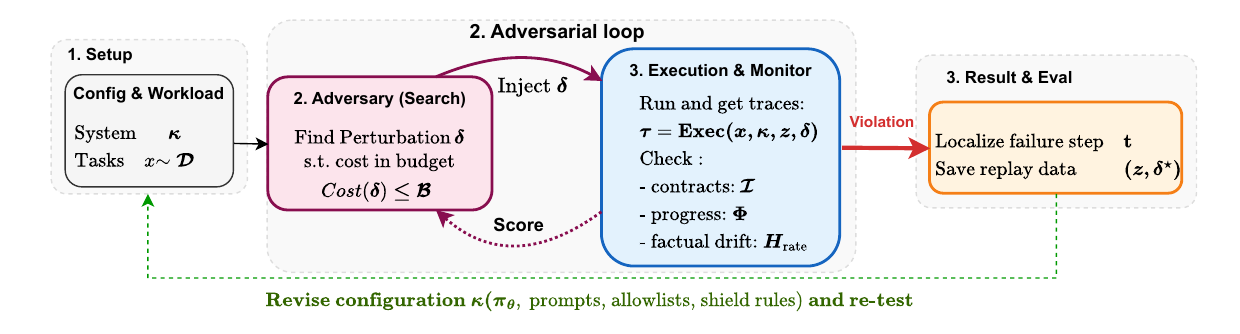}
\caption{\textbf{Adversarial counterexample search as an inner--outer assurance loop.}
\textbf{(1) Setup:} fix a system configuration $\kappa$ (roles, tools, contracts, governance) and sample tasks $x \sim \mathcal{D}$ with stochastic seed $z$.
\textbf{(2) Inner loop (search):} an adversary selects bounded perturbations $\delta$ (subject to $\mathrm{cost}(\delta)\le B$) and injects them into the execution; the system produces a trace $\tau=\mathrm{Exec}(x,\kappa,z,\delta)$, which is monitored for contract violations (step/trace contracts $\mathcal{I}^{\mathrm{step}},\mathcal{I}^{\mathrm{trace}}$) and auxiliary signals such as progress $\Phi$ and unsupported-claim rate $H_{\mathrm{rate}}$. The resulting score guides the next perturbation choice.
\textbf{(3) Outer loop (engineering feedback):} when a violation is found (red arrow), the framework localizes the first failing step $t$ and stores a replay record (e.g., $(z,\delta^\star)$ and required stubs), enabling configuration revision and re-testing (green dashed path), including updates to agent parameters $\pi_\theta$ and/or governance policy $\Pi$.}
\label{fig:adversarial_loop}
\end{figure*}

\noindent\textit{Search strategies and adaptive adversaries.}
The same perturbation operator interface can be used with simple search methods (random fuzzing, beam search over operator sequences, evolutionary search), reinforcement learning, or with an adaptive adversary.

Adaptive search is useful for staged failures that unfold over multiple steps (e.g., first inducing role drift, then exploiting the resulting authority).
At step $t$, the adversary observes a filtered summary $h_t=\mathrm{Obs}(r_t)$ (exposing only safe MAT fields such as action type, called service, and contract pass/fail outcomes) and selects the next operator $\delta_t$.
Figure~\ref{fig:adversarial_loop} summarizes this inner-loop interaction.

When the adversary is implemented as a learned policy (e.g., via reinforcement learning), the inner-loop perturbation selection can be formalized as an optimization problem.
The adversary observes a filtered state summary $h_t=\mathrm{Obs}(r_t)$ and selects perturbation operators $\delta_t$ to maximize contract violations while remaining within a bounded cost budget. 
Equation~(\ref{eq:adv-objective}) defines the objective optimized by such an adaptive adversary. Here, $\mathrm{Instab}(x,\kappa,\delta)$ penalizes perturbation schedules that lead to unstable or non-reproducible executions (e.g., high variance across replays), encouraging replayable counterexamples.

\begin{equation}
\label{eq:adv-objective}
\begin{alignedat}{2}
\max_{\pi_{\psi}^{\mathrm{adv}}}\quad &
\mathbb{E}\!\Bigg[
\sum_{t=0}^{T} \gamma^t
\Big(
\mathrm{Vio}(r_t,\mathcal{I}^{\mathrm{step}}_t)
-\beta\,\mathrm{cost}(\delta_t)
\Big)
\Bigg] \\
&\;-\lambda\,\mathbb{E}\!\Big[\mathrm{Instab}(x,\kappa,\delta)\Big] \\
\text{s.t.}\quad &
\sum_{t=0}^{T} \mathrm{cost}(\delta_t) \le B .
\end{alignedat}
\end{equation}

\noindent\textit{Operator families.}
The operator space $\Delta$ is a small library of bounded transformations:
\begin{itemize}
    \item \emph{Prompt and context:} paraphrasing, adding ambiguity, reordering constraints, inserting distractors, or splicing untrusted text, bounded by token changes.
    \item \emph{External services:} delays and timeouts, dropped responses, field corruption, metadata changes, or stale results, bounded by frequency and magnitude.
    \item \emph{Retrieval:} shuffling top $k$, adding near duplicate distractors, perturbing ranking scores, or truncating evidence, bounded by the number of modified items.
    \item \emph{Memory:} inserting untrusted notes, reordering items, forcing truncation, or simulating write collisions, bounded by the number and size of writes.
\end{itemize}

\noindent\textit{From failures to reproducible test-cases.}
For each stress testing experiment (one executed search on a task and system configuration), the procedure returns a replayable perturbation schedule $\delta^\star_{0:T}$ and the violated contracts with the responsible steps $(t,i_t)$ identified from MAT.
In addition, a replay record is stored, including the seed, tool stubs or cached tool outputs, retrieved artifacts, and the injected fault schedule. This allows deterministic reproduction in continuous integration and supports regression tests after changes to prompts, policies, or services. The replay artifact supports regression testing after changes to $\kappa$, including updates to the agent policy $\pi_\theta$ and adjustments to governance policy $\Pi$ (Figure~\ref{fig:adversarial_loop}, green path).

\subsection{L3: Structured fault injection across external services, retrieval, and memory} \label{sec:l3}
Adversarial perturbations are only one source of failures.
Production assurance also requires robustness to common integration faults such as timeouts, partial outages, stale caches, malformed payloads, and inconsistent shared memory.
A chaos engineering \cite{chaos_engineering} approach is used to test these conditions by injecting controlled faults at system boundaries.
During the run, MAT contracts check a \emph{containment} property.
Containment requires that the fault is detected, an appropriate mitigation is executed, and the final user-facing output still satisfies the required contracts.

\noindent\textit{Fault operators and injection points.}
Faults are injected at three boundaries: service adapters, retrieval, and shared memory. The fault library includes timeouts and delays, dropped or partial responses, payload corruption, schema mismatches, stale cache returns, retrieval shuffles, and memory injection, reordering, or collisions. Parameters such as $\ell$, $\eta$, and $\Delta t$ bound fault severity.

\noindent\textit{Containment contract.}
Let $\mathrm{Fault}_t$ indicate that a fault is injected or observed at step $t$, $\mathrm{Detect}_t$ that the fault is recognized and recorded in MAT, and $\mathrm{Mitigate}_t$ that a corrective action is taken (retry and backoff, alternate service, replan, clarification, or escalation). A minimal containment requirement is:
\begin{equation}
\label{eq:containment}
\begin{aligned}
\forall\, t:\ \mathrm{Fault}_t \Rightarrow \ &\exists\, t' \in [t, t+w]\ \mathrm{Detect}_{t'}\\
&\wedge\ \exists\, t'' \in [t', t'+w']\ \mathrm{Mitigate}_{t''}.
\end{aligned}
\end{equation}
In addition, the final response must satisfy all relevant contracts, including all $I^{\mathrm{trace}}_k\in\mathcal{I}^{\mathrm{trace}}$ and any end of run $I^{\mathrm{step}}_k$ checks that apply to the final response. If Eq.~\ref{eq:containment} fails, MAT localizes the boundary where the fault entered and the first step where it led to a trace-level violation, enabling replay and regression testing (under recorded seeds and stubs) using the recorded seed and fault schedule.

\subsection{L4: Governing external actions and service calls}\label{sec:l4}
Even when a system performs well on routine tasks, delegated tool access can turn small reasoning errors or injected instructions into real-world harm. Governance is therefore treated as a runtime layer that mediates external actions via least privilege, policy enforcement, and selective escalation for high-impact effects.

\noindent\textit{Capability boundaries (least privilege).}
Each agent $A_i$ (for $i \in \mathcal{A}$) is assigned a capability set $\mathcal{K}_i$ (allowed services, endpoints, parameter ranges, and rate limits).
Let $\mathrm{cap}(a_t)$ denote the capabilities required by action $a_t$.
Action $a_t$ is authorized for the acting agent $i_t$ iff:
\[
\mathrm{Allow}(a_t,i_t)=1 \iff \mathrm{cap}(a_t)\subseteq \mathcal{K}_{i_t}.
\]
This limits the impact of prompt injection and behavioral drift by construction.

\noindent\textit{Policy shield (allow / rewrite / block).}
All proposed actions pass through a policy mediator $\Pi$, which produces a governed outcome given the proposed action $a_t$ and the current state estimate $\hat{s}_t$:
\begin{equation}
\label{eq:policy-shield}
\Pi(a_t,\hat{s}_t)\in\{\op{allow}(a_t),\ \op{rewrite}(\tilde a_t),\ \op{block}\}.
\end{equation}
Rewriting applies bounded corrections (e.g., removing sensitive content, clamping parameters to safe ranges, or converting a high-impact action into an approval request). Approval can be provided by a human reviewer (human-in-the-loop, HITL) or by a designated verifier component, depending on deployment policy.

\noindent\textit{Risk-aware routing and sandboxing (trust as a heuristic).}
A per-agent trust score $T_i(t)\in[0,1]$ summarizes recent contract outcomes and policy events. Trust is not a security guarantee; it is a practical signal used for routing and sandboxing, with decay so older ``good'' behavior does not dominate. When $T_i(t)$ falls below a threshold, governance can: (i) shrink the capability set $\mathcal{K}_i$, (ii) route the subtask to a verifier agent, or (iii) apply stricter mediation for high-impact actions.
\noindent\textit{Selective human escalation for high-impact actions.}
For pre-labeled high-impact actions (e.g., \op{send\_email}, \op{payment}, \op{delete}, \op{grant\_access}), policy may require HITL approval. Operationally, this requirement is enforced by the policy mediator $\Pi$ via \op{rewrite} (convert the action into an explicit approval request) or \op{block} (prevent execution until approval), as defined in Eq.~(\ref{eq:policy-shield}).

\section{Evaluation and Metrics}
\label{sec:eval}

This section proposes an evaluation methodology and a set of metrics for assessing multi-agent LLM systems and governance variants.
The methodology treats evaluation as contract-monitored execution traces, so that utility, robustness to perturbations, fault containment, and governance behavior are comparable under a common trace representation.
A full empirical study is left to ongoing and future work; the protocol and estimators below are specified in a form intended to be instantiated in a later implementation.

\subsection{Evaluation object: tasks, configurations, and traces}
Let $x \sim \mathcal{D}$ denote a task instance sampled from a workload distribution.
Let $\kappa$ denote a system configuration (agent topology, role prompts, tool adapters, contract set, and governance policy).
Let $z$ denote a stochastic seed (model sampling, tie breaks, and any randomized tool stubs).
Let $\delta$ denote a perturbation and fault schedule drawn from an operator family $\Delta$.
A single run produces an execution trace:
\[
\tau \;=\; \mathrm{Exec}(x,\kappa,z,\delta).
\]
A run is a \emph{contract failure} if at least one trace contract is violated, using Eq.~(\ref{eq:fail-def}).
A run is a \emph{task failure} if the task objective is not met, even if no contract is violated.
Both notions are reported because they reflect different deployment concerns.
Contract failure is safety and governance oriented: it indicates that at least one monitored constraint (e.g., verification gates, least privilege, containment requirements, refusal rules) has been violated, even if the task output appears useful.
Task failure is end-to-end utility oriented: it indicates that the intended outcome was not achieved (wrong, incomplete, or unusable), even if monitored constraints were satisfied.

To estimate expectations, a finite workload set $\mathcal{D}=\{x_1,\dots,x_N\}$ is fixed and each task is run with $S$ seeds $z_1,\dots,z_S$, yielding $N\!\times\!S$ runs per $(\kappa,\delta)$ condition.
Rate metrics that reduce to a binomial proportion (e.g., success, contract failure, containment, non-termination, refusal or block rates) are accompanied by Wilson-score confidence intervals or Clopper--Pearson exact intervals.
For derived metrics that are not simple proportions (e.g., ratios such as tokens per successful run, or robustness summaries such as area under $R(B)$), percentile bootstrap intervals are computed by resampling runs and recomputing the metric.

\subsection{Protocol and design choices}
Industry guidance has emphasized the practical importance of well-scoped task suites and systematic failure analysis when evaluating agentic systems\footnote{\url{https://www.anthropic.com/engineering/demystifying-evals-for-ai-agents}}. The protocol specified here complements these recommendations with contract-monitored traces that localize failures to concrete execution steps and interfaces.
\noindent\textit{Workloads.}
An enterprise-oriented workload suite $\mathcal{D}$ is assumed to cover common orchestration patterns and risk cases.
A practical instantiation partitions tasks into categories and uses a fixed number of tasks per category, so that aggregate results are not dominated by a single class:
\begin{itemize}
    \item \textit{Tool use tasks:} query an internal database, retrieve documents, summarize, and propose an action.
    \item \textit{Multi-step planning tasks:} delegation across roles, approval gates, retries, and coordination across agents.
    \item \textit{Policy constrained tasks:} PII handling, consent lists, and compliance checks enforced at action time.
    \item \textit{Misuse oriented tasks:} malicious or policy violating requests that should trigger refusal or containment.
\end{itemize}
Each task includes at least one untrusted input channel (e.g., email text, a retrieved snippet, or a document excerpt) and at least one tool boundary where integration failures are realistic (e.g., timeouts, partial results, stale cache returns, or schema mismatch).
This supports direct measurement of injection robustness, fault containment, and governance behavior.

Existing agent evaluation suites are used to seed task specifications and prompt templates (e.g., AgentBench \cite{agentbench}, GAIA \cite{gaia}, AgentDojo \cite{agentdojo}, HarmBench \cite{harmbench}, AgentHarm \cite{agentharm}).
These suites provide realistic task narratives and failure oriented scenarios, but they typically assume their own tool abstractions, action schemas, and safety rules.
For trace contract evaluation, each seeded task is re-instantiated for the target deployment by adapting (a) the tool layer (swap benchmark tools for the deployed adapters and endpoints), (b) the action interface (map benchmark actions onto the system action types: messages, tool calls, memory reads and writes, delegation, termination), and (c) the policy layer (align PII, consent, and capability constraints with the governance policy).
In addition, the task is paired with a perturbation and fault matrix applied at the same tool and memory boundaries that are monitored by contracts.
This instantiation procedure provides a practical bridge from existing task suites to contract-monitored, tool grounded evaluations of agent interactions.

\noindent\textit{Perturbation and fault matrix.}
For each task instance $x$, evaluation covers three operating conditions:
\begin{itemize}
    \item \textit{Nominal runs ($\delta=\varnothing$):} baseline behavior under the unmodified input and normal tool operation. These runs measure utility and contract compliance without injected disturbances.
    \item \textit{Structured fault injection ($\delta \in \Delta_{\mathrm{fault}}$):} controlled boundary faults applied to tools, retrieval, or memory. These runs measure robustness to integration issues such as timeouts, partial responses, stale cache returns, and schema mismatches.
    \item \textit{Adversarial schedules ($\delta$ chosen by search under a cost budget):} perturbations selected to trigger contract violations while remaining small and realistic. These runs measure worst-case behavior under bounded, plausible disturbances.
\end{itemize}
This design reflects the observation that production reliability depends on behavior under imperfect conditions, not only under nominal inputs \cite{chaos_engineering,pan_measuring_agents_prod}.

\noindent\textit{Governance variants and ablations.}
Results are compared across governance and instrumentation variants. A minimal set includes:
\begin{itemize}
    \item \textit{No governance.} Tool actions are executed directly, without policy mediation.
    \item \textit{Policy shield.} Each proposed tool action is mediated at execution time (allow / rewrite / block), as in Eq.~\ref{eq:policy-shield}.
    \item \textit{Shield + capability constraints.} Least privilege is enforced via per-agent capability sets $\mathcal{K}_i$, which restrict the services, endpoints, and parameter ranges that each agent may invoke.
    \item \textit{Shield + routing heuristics.} Risk-aware routing and sandboxing are enabled; when recent contract outcomes indicate elevated risk, actions are routed through stricter mediation or forwarded to a verifier or human approval step.
\end{itemize}
These variants target risks emphasized by OWASP and related guidance, including prompt injection and excessive agency \cite{owasp_llm_top10,mcp_security_best_practices,unit42_mcp_attack_vectors}.

\noindent\textit{Repeated runs and seed control.}
Each task is run with multiple seeds.
This captures stochastic variability from model sampling and tool or retrieval non-determinism.
Single-run metrics are reported alongside multi-run robustness metrics (e.g., Success@k).
For failure analysis, an execution record for deterministic reproduction is stored, including seeds, tool stubs, and injected fault schedules; this enables later regression testing against the same recorded conditions.

\noindent\textit{Uncertainty quantification.}
For binomial proportions (success rates, violation rates, containment rates), Wilson-score confidence intervals or Clopper--Pearson exact intervals are reported \cite{Antwi2025AdaptiveBayesianRareBinomial}.
For non-linear aggregates (e.g., robustness curve area, tokens per successful run), bootstrap confidence intervals are reported \cite{efron_tibshirani_bootstrap}.
Per-category results (means and intervals) are also reported to avoid mixing heterogeneous task types.

\subsection{Primary metrics}
Metrics are grouped by evaluation goal. All metrics are computed from MAT traces and monitored contracts.
For estimation, fix a finite workload set $\mathcal{D}=\{x_1,\dots,x_N\}$ and $S$ seeds per task, $z_1,\dots,z_S$ (so there are $N\!\times\!S$ total runs).
For a fixed configuration $\kappa$ and operating condition $\delta$, run $(x_i,z_s)$ produces a trace $\tau_{i,s}=\mathrm{Exec}(x_i,\kappa,z_s,\delta)$ and an outcome indicator $Y(x_i,z_s,\delta)\in\{0,1\}$.

\subsubsection{End-to-end utility and reliability}
\paragraph{Task success}
Task success measures whether the task objective is satisfied for a given run.
The estimated success rate under configuration $\kappa$ and condition $\delta$ is:
\[
\widehat{\mathrm{Success}}(\kappa,\delta)
\;=\;
\frac{1}{NS}\sum_{i=1}^{N}\sum_{s=1}^{S} Y(x_i,z_s,\delta).
\]

\paragraph{Success@k}
Because the system is stochastic, a robustness metric analogous to pass@k is defined and is reported when multi-run evaluation is included \cite{chen2021_codex}.
A task is counted as solved if any of $k$ independent runs succeeds:
\[
\widehat{\mathrm{Success@k}}(\kappa,\delta)
\;=\;
\frac{1}{N}\sum_{i=1}^{N}\mathbb{I}\left\{\max_{j\in\{1,\dots,k\}} Y(x_i,z_j,\delta)=1\right\}.
\]
This metric is reported for nominal runs and, when robustness is evaluated, for selected perturbation budgets.

\paragraph{Termination reliability}
Non-termination and coordination collapse are measured using $\mathrm{Term}(\tau)$ and the progress contract (Eq.~\ref{eq:coord-collapse}).
The estimated non-termination rate is:
\[
\widehat{\mathrm{NTR}}
\;=\;
\frac{1}{NS}\sum_{i=1}^{N}\sum_{s=1}^{S}\mathbb{I}\{\neg \mathrm{Term}(\tau_{i,s})\}.
\]
To distinguish early failures from long-horizon collapse, distributions of steps-to-termination and steps-to-first-failure can be included.

\subsubsection{Contract compliance and failure localization}
\paragraph{Trace contract violation rate}
Trace-level assurance is summarized by the fraction of runs that violate at least one trace contract (Eq.~\ref{eq:fail-def}):
\[
\widehat{\mathrm{Fail}}
\;=\;
\frac{1}{NS}\sum_{i=1}^{N}\sum_{s=1}^{S}
\mathrm{Fail}(\tau_{i,s},\mathcal{I}^{\mathrm{trace}}).
\]

\paragraph{Per-contract violation profile}
Let $\mathrm{VioID}(\tau)$ denote the set of violated contract IDs recorded in MAT verdicts.
For each contract $k$, the estimated violation rate is:
\[
\widehat{p}_k
\;=\;
\frac{1}{NS}\sum_{i=1}^{N}\sum_{s=1}^{S}\mathbb{I}\{k \in \mathrm{VioID}(\tau_{i,s})\}.
\]
A ranked list of contracts by $\widehat{p}_k$ is reported.
Hard and soft violations are separated using the severity field in Eq.~(\ref{eq:verdict}).

\paragraph{First violation step and responsible role}
For each failing run, let $t^\star$ be the first step where any monitored contract is violated.
The distributions of $t^\star$ and the associated agent identity $i_{t^\star}$ are reported.
This indicates whether failures tend to arise early (planning and delegation) or late (tool actions, policy mediation, or escalation).

\subsubsection{Quality and factuality}
\paragraph{Unsupported claim rate and propagation}
The definitions from failure class F2 (Section~\ref{sec:taxonomy}) are reused.
Given claims $\{c_j\}$ extracted from the final response and provenance evidence $\{e_j\}$ recorded in MAT, $H_{\mathrm{rate}}$ and $H_{\mathrm{prop}}$ are computed per run and then averaged across runs:
\[
\widehat{H}_{\mathrm{rate}} \;=\; \frac{1}{NS}\sum_{i,s} H_{\mathrm{rate}}(\tau_{i,s}),\;
\widehat{H}_{\mathrm{prop}} \;=\; \frac{1}{NS}\sum_{i,s} H_{\mathrm{prop}}(\tau_{i,s}).
\]
When gold reference answers exist, support can be checked by exact match or entailment tests.
When gold is unavailable, a constrained judge may be used; in that case, judge calibration and judge variance are reported \cite{zheng2023_mtbench}.

\paragraph{Retrieval-grounded metrics}
For configurations that include a retrieval component (i.e., the system consults an external corpus and conditions generation on retrieved passages), reference-free retrieval quality metrics such as faithfulness and context relevance are reported using established tooling \cite{ragas}.
These metrics are treated as secondary signals, mainly to interpret changes in unsupported-claim rates under retrieval noise.

\subsubsection{Fault robustness and containment}
\paragraph{Containment rate}
For structured fault injection runs, the containment rate (Sec.~\ref{sec:framework}, Eq.~\ref{eq:containment}) reports how often injected faults are detected and mitigated within the required windows and do not lead to trace-level failures in the final output: 
\[
\widehat{\mathrm{CR}}
\;=\;
\frac{\#\{\text{faults that satisfy Eq.~(\ref{eq:containment}) and final contracts}\}}
{\#\{\text{faults injected}\}}.
\]
A breakdown by fault type can also be included (timeout, partial response, stale cache, corrupt payload, memory collision).

\paragraph{Residual harm under faults}
Containment can fail by delayed detection, ineffective mitigation, or downstream policy violations.
An additional safety signal is therefore reported: whether any policy-relevant contract is violated in the final response after a fault.
This complements $\widehat{\mathrm{CR}}$ by separating non-critical quality loss from safety-critical failure.

\subsubsection{Safety, governance, and misuse resistance}
\paragraph{Policy mediator outcomes}
The governance layer produces allow, rewrite, or block outcomes (Eq.~\ref{eq:policy-shield}).
Their occurrence rates over tool actions are reported:
\[
\widehat{p}_{\mathrm{allow}},\quad \widehat{p}_{\mathrm{rewrite}},\quad \widehat{p}_{\mathrm{block}}.
\]
Rates are also reported conditioned on action risk level (low-impact vs.\ high-impact) and input category (non-malicious vs.\ misuse-oriented).

\paragraph{Blocked high-impact action rate}
For pre-labeled high-impact actions, the fraction that are blocked or routed to approval is reported.
This provides a measurable signal of excessive agency at the action boundary.

\paragraph{Misuse success and refusal}
For misuse-oriented tasks, three outcome rates are reported:
\begin{itemize}
    \item refusal or safe abort,
    \item harmful completion,
    \item partial completion with containment (side effects blocked).
\end{itemize}
These outcomes are defined using task labels and policy contracts \cite{agentharm,harmbench}.

\subsubsection{Efficiency and operational cost}
\paragraph{Token and tool cost per successful task}
Tokens per successful run are reported as:
\[
\widehat{T}_{\mathrm{eff}}
\;=\;
\frac{\sum_{i,s} \text{tokens}(\tau_{i,s})}{\#\{(i,s): Y(x_i,z_s,\delta)=1\}}.
\]
Tool-call counts and latency summaries are reported separately.
Efficiency is not treated as a primary objective, but is included to contextualize overhead introduced by governance and verification \cite{pan_measuring_agents_prod}.

\subsection{Robustness, MTBF, and maintainability}
Beyond nominal performance, evaluation may also summarize robustness under bounded perturbations, stability under continuous operation, and regressions after system changes.

\paragraph{Robustness under perturbation budgets}
Robustness is defined as a function of a perturbation budget.
Let $B$ be a maximum budget on perturbation cost (Sec.~\ref{sec:framework}).
Define the robustness curve as the expected task success when the perturbation schedule is chosen within budget $B$:
\[
R(B)
\;=\;
\frac{1}{NS}\sum_{i=1}^{N}\sum_{s=1}^{S} Y\!\big(x_i,z_s,\delta^\star(B)\big),
\]
where $\delta^\star(B)$ denotes a perturbation schedule selected under the budget constraint $\mathrm{cost}(\delta)\le B$ (e.g., by sampling schedules from $\Delta$ subject to the cost budget, or by an adaptive search procedure).
The curve $R(B)$ may be reported for a small set of budget values.
A compact summary may also be reported, such as the area under $R(B)$ over a fixed budget range, together with the success rate at the largest tested budget.
Results may be reported per workload category to avoid masking brittle behavior on risk-heavy tasks.

\paragraph{Mean time between failures (MTBF)}
MTBF is most meaningful when the system processes a stream of tasks or operates continuously.
An episode is defined as a sequence of tasks processed under the same configuration, with state carried across tasks as appropriate (e.g., persistent memory, caches, and trust scores).
Let $F_r$ be the number of MAT steps until the first trace-contract failure in episode $r$, or until the episode ends.
The estimator is:
\[
\widehat{\mathrm{MTBF}}
\;=\;
\frac{1}{R}\sum_{r=1}^{R} F_r.
\]
MTBF may be reported in steps and in wall-clock time when latency measurements are available.

\paragraph{Regression and maintainability}
Prompt, tool, and policy changes are frequent in deployed systems.
This motivates regression testing using stored execution artifacts (e.g., seeds and injected fault schedules) so that failures can be reproduced after changes.
Let $\mathrm{Pass}(x,\kappa)=1$ denote that task $x$ completes successfully and satisfies all required contracts under configuration $\kappa$ under nominal conditions and, optionally, under a fixed perturbation set.
Given a regression suite $\mathcal{D}_{\mathrm{reg}}$ and configurations $\kappa$ (old) and $\kappa'$ (new), define the regression rate as:
\[
\widehat{R}_{\mathrm{reg}}
\;=\;
\frac{\#\{x \in \mathcal{D}_{\mathrm{reg}}:\ \mathrm{Pass}(x,\kappa)=1 \wedge \mathrm{Pass}(x,\kappa')=0\}}
{\#\{x \in \mathcal{D}_{\mathrm{reg}}:\ \mathrm{Pass}(x,\kappa)=1\}}.
\]
Regressions are reported separately for task success, contract compliance, and governance outcomes (allow vs.\ rewrite vs.\ block changes), since these capture different operational risks.

\paragraph{Coverage units and regression-suite selection} The replay artifacts produced by contract violations can be treated as regression test cases.
Unlike code coverage, the framework supports \emph{contract- and interface-oriented} coverage.
Each stored replay is indexed by a \emph{failure signature} derived from MAT, including (i) the first violated contract ID and severity, (ii) the triggering action type (e.g., tool call, memory write, delegation, final response), (iii) the affected interface class (service, retrieval, memory), (iv) the action risk label (high-impact vs.\ low-impact), and (v) the perturbation or fault operator family.
A compact regression suite can then be obtained by selecting a minimal subset of replays that covers all observed high-severity contracts (and representative signatures per interface and action type), together with a small set of nominal passing tasks per workload category to detect quality regressions. Formally, suite selection can be posed as a set-cover problem over these coverage units; however, the framework is independent of the particular solver used.

\subsection{Recommended reporting format}
A consistent reporting structure improves interpretability and makes comparisons across governance variants reproducible.
The following artifacts are typically sufficient.

\noindent\textit{Nominal performance table.}
Recommended reporting under nominal conditions ($\delta=\varnothing$) includes, for each governance variant: end-to-end task success (and Success@k when stochasticity is material), non-termination or coordination-collapse rate, and basic efficiency signals (steps, tool calls, and tokens per successful run).
For mixed task suites, per-category results are recommended to avoid masking failures on harder categories.

\noindent\textit{Fault-injection outcomes table.}
Report results under structured fault injection ($\delta\in\Delta_{\mathrm{fault}}$) in a separate table.
Summarize \emph{containment} in simple rate form:
how often an injected fault is: (a) detected within the next $w$ steps, and then (b) followed by a mitigation within the next $w'$ steps.
It may also be useful to report whether the final response satisfies the relevant trace contracts after mitigation.
For systems with action mediation, include the distribution of policy outcomes (allow, rewrite, block) to distinguish recovery from conservative blocking.

\noindent\textit{Robustness curves.}
Plot robustness curves $R(B)$ as a function of perturbation budget $B$ for at least two workload categories.
Where applicable, include both random perturbations and adversarial schedules under the same budget.
Optionally add a scalar summary such as the area under the robustness curve over a fixed budget range.

\noindent\textit{Contract violation profile and examples.}
A compact contract violation profile may be included as a ranked list of the most frequently violated contracts.
Alongside frequency, report a localization statistic such as the most common first-violation step index or the dominant action type at first violation.
Optionally add a small number of short, reproducible counterexamples.
Each example should identify the triggering perturbation or injected fault, the first violated contract, and the localized MAT step window around the violation.

\section{Conclusion}

This paper addressed assurance for agentic systems in which one or more orchestrators coordinate multiple agents that interact with external services, retrieval components, and shared memory. In this setting, failures are not limited to incorrect final responses; they also include non-termination, coordination breakdowns, error propagation across agents, role drift, and unsafe interface-driven actions.

We presented a trace-based assurance framework that organizes evaluation around contract-monitored executions. The framework (i) instruments runs as Message-Action Traces with step- and trace-level contracts, enabling localization of the first violating step and replay; (ii) formulates stress testing as budgeted counterexample search over bounded perturbations; (iii) supports structured fault injection at service, retrieval, and memory interfaces to test containment under realistic integration disturbances; and (iv) treats governance as a runtime component that mediates external actions through capability restrictions and allow/rewrite/block decisions at the language-to-action boundary. We also defined trace-derived metrics that support comparison across stochastic seeds, models, and orchestration configurations.

\noindent\textit{Limitations and next steps.}
The current contribution focuses on methodology, definitions, and evaluation estimators rather than an empirical demonstration. The next step is to implement the instrumentation, contract library, perturbation and fault operators, and governance mechanisms in a prototype, and to execute a systematic empirical study across representative workloads and orchestration variants. This study will quantify utility, robustness, containment, and governance behavior, and will assess reproducibility via replay artifacts and regression testing.

Future work will also study coverage criteria and selection strategies for maintaining small but effective regression suites, including contract coverage, boundary/interface coverage, and risk-conditioned coverage.
Another direction is integrating the protocol into the software development lifecycle as evaluation-driven development: changes to configuration $\kappa$ (prompts, tool adapters, allowlists), governance policy $\Pi$, or agent parameters $\pi_\theta$ would be accepted only if a fixed smoke suite and a replay-based regression suite preserve contract compliance, containment, and utility within predefined statistical tolerances across seeds.

\section*{Acknowledgments}
This work was supported by the project “Romanian Hub for Artificial Intelligence (HRIA)”, funded under the Smart Growth, Digitization and Financial Instruments Programme 2021--2027, MySMIS no. 351416. We also thank Adobe Romania for funding a PhD scholarship on the topic and for fruitful discussions and feedback that helped shape this work.

\bibliographystyle{IEEEtran}
\bibliography{references}

\begin{thebibliography}{10}
\providecommand{\url}[1]{#1}
\csname url@samestyle\endcsname
\providecommand{\newblock}{\relax}
\providecommand{\bibinfo}[2]{#2}
\providecommand{\BIBentrySTDinterwordspacing}{\spaceskip=0pt\relax}
\providecommand{\BIBentryALTinterwordstretchfactor}{4}
\providecommand{\BIBentryALTinterwordspacing}{\spaceskip=\fontdimen2\font plus
\BIBentryALTinterwordstretchfactor\fontdimen3\font minus
  \fontdimen4\font\relax}
\providecommand{\BIBforeignlanguage}[2]{{%
\expandafter\ifx\csname l@#1\endcsname\relax
\typeout{** WARNING: IEEEtran.bst: No hyphenation pattern has been}%
\typeout{** loaded for the language `#1'. Using the pattern for}%
\typeout{** the default language instead.}%
\else
\language=\csname l@#1\endcsname
\fi
#2}}
\providecommand{\BIBdecl}{\relax}
\BIBdecl

\bibitem{pan_measuring_agents_prod}
\BIBentryALTinterwordspacing
M.~Z. Pan \emph{et~al.}, ``Measuring agents in production,'' arXiv, 2025.
  [Online]. Available: \url{https://arxiv.org/abs/2512.04123}
\BIBentrySTDinterwordspacing

\bibitem{wef_capgemini_agents}
\BIBentryALTinterwordspacing
{World Economic Forum} and {Capgemini}, ``Ai agents in action: Foundations for
  evaluation and governance,'' White Paper, Nov. 2025, accessed: 2025-12-17.
  [Online]. Available:
  \url{https://www.weforum.org/publications/ai-agents-in-action-foundations-for-evaluation-and-governance/}
\BIBentrySTDinterwordspacing

\bibitem{rv_survey_sanchez}
\BIBentryALTinterwordspacing
C.~Sanchez \emph{et~al.}, ``A survey of challenges for runtime verification
  from advanced application domains,'' \emph{Formal Methods in System Design},
  2018, preprint: arXiv:1811.06740. [Online]. Available:
  \url{https://arxiv.org/abs/1811.06740}
\BIBentrySTDinterwordspacing

\bibitem{bauer2011rv}
A.~Bauer, M.~Leucker, and C.~Schallhart, ``Runtime verification for {LTL} and
  {TLTL},'' \emph{ACM Transactions on Software Engineering and Methodology},
  vol.~20, no.~4, pp. 14:1--14:64, 2011.

\bibitem{chaos_engineering}
\BIBentryALTinterwordspacing
A.~Basiri \emph{et~al.}, ``Chaos engineering,'' \emph{IEEE Software}, vol.~33,
  no.~3, pp. 35--41, 2016, also available as arXiv:1702.05843. [Online].
  Available: \url{https://arxiv.org/abs/1702.05843}
\BIBentrySTDinterwordspacing

\bibitem{owasp_llm_top10}
\BIBentryALTinterwordspacing
{OWASP Foundation}, ``Owasp top 10 for large language model applications,''
  OWASP Project Page, 2025, accessed: 2025-12-17. [Online]. Available:
  \url{https://owasp.org/www-project-top-10-for-large-language-model-applications/}
\BIBentrySTDinterwordspacing

\bibitem{mcp_security_best_practices}
\BIBentryALTinterwordspacing
{Model Context Protocol (MCP)}, ``Security best practices (draft),'' Online
  documentation, 2025, accessed: 2025-12-17. [Online]. Available:
  \url{https://modelcontextprotocol.io/specification/draft/basic/security_best_practices}
\BIBentrySTDinterwordspacing

\bibitem{unit42_mcp_attack_vectors}
\BIBentryALTinterwordspacing
{Palo Alto Networks Unit 42}, ``New prompt injection attack vectors through
  {MCP} sampling,'' Blog post, Dec. 2025, accessed: 2025-12-17. [Online].
  Available:
  \url{https://unit42.paloaltonetworks.com/model-context-protocol-attack-vectors/}
\BIBentrySTDinterwordspacing

\bibitem{HECKEL2005145}
R.~Heckel and M.~Lohmann, ``Towards contract-based testing of web services,''
  \emph{Electronic Notes in Theoretical Computer Science}, vol. 116, pp.
  145--156, 2005, tACoS.

\bibitem{autogen}
\BIBentryALTinterwordspacing
Q.~Wu \emph{et~al.}, ``{AutoGen}: Enabling next-gen {LLM} applications via
  multi-agent conversation framework,'' OpenReview, 2023, accessed: 2025-12-17.
  [Online]. Available: \url{https://openreview.net/forum?id=BAakY1hNKS}
\BIBentrySTDinterwordspacing

\bibitem{langgraph}
\BIBentryALTinterwordspacing
{LangChain}, ``{LangGraph} documentation (overview),'' Online documentation,
  2025, accessed: 2025-12-17. [Online]. Available:
  \url{https://docs.langchain.com/oss/python/langgraph/overview}
\BIBentrySTDinterwordspacing

\bibitem{langsmith}
\BIBentryALTinterwordspacing
------, ``{LangSmith} evaluation documentation,'' Online documentation, 2025,
  accessed: 2025-12-17. [Online]. Available:
  \url{https://docs.langchain.com/langsmith/evaluation}
\BIBentrySTDinterwordspacing

\bibitem{ragas}
\BIBentryALTinterwordspacing
S.~Es \emph{et~al.}, ``{RAGAS}: Automated evaluation of retrieval augmented
  generation,'' arXiv, 2023. [Online]. Available:
  \url{https://arxiv.org/abs/2309.15217}
\BIBentrySTDinterwordspacing

\bibitem{agentbench}
\BIBentryALTinterwordspacing
X.~Liu \emph{et~al.}, ``Agentbench: Evaluating {LLMs} as agents,'' arXiv, 2023.
  [Online]. Available: \url{https://arxiv.org/abs/2308.03688}
\BIBentrySTDinterwordspacing

\bibitem{gaia}
\BIBentryALTinterwordspacing
G.~Mialon \emph{et~al.}, ``{GAIA}: A benchmark for general {AI} assistants,''
  arXiv, 2023. [Online]. Available: \url{https://arxiv.org/abs/2311.12983}
\BIBentrySTDinterwordspacing

\bibitem{bipia}
\BIBentryALTinterwordspacing
J.~Yi \emph{et~al.}, ``Benchmarking and defending against indirect prompt
  injection attacks on large language models,'' arXiv, 2023. [Online].
  Available: \url{https://arxiv.org/abs/2312.14197}
\BIBentrySTDinterwordspacing

\bibitem{agentdojo}
\BIBentryALTinterwordspacing
E.~Debenedetti \emph{et~al.}, ``Agentdojo: A dynamic environment to evaluate
  attacks and defenses for {LLM} agents,'' arXiv, 2024. [Online]. Available:
  \url{https://arxiv.org/abs/2406.13352}
\BIBentrySTDinterwordspacing

\bibitem{wasp}
\BIBentryALTinterwordspacing
I.~Evtimov \emph{et~al.}, ``{WASP}: Benchmarking web agent security against
  prompt injection attacks,'' arXiv, 2025. [Online]. Available:
  \url{https://arxiv.org/abs/2504.18575}
\BIBentrySTDinterwordspacing

\bibitem{injecagent_acl}
\BIBentryALTinterwordspacing
Q.~Zhan \emph{et~al.}, ``Injecagent: Benchmarking indirect prompt injections in
  tool-integrated large language model agents,'' arXiv (Findings of ACL 2024),
  2024. [Online]. Available: \url{https://arxiv.org/abs/2403.02691}
\BIBentrySTDinterwordspacing

\bibitem{agentharm}
\BIBentryALTinterwordspacing
M.~Andriushchenko \emph{et~al.}, ``Agentharm: A benchmark for measuring
  harmfulness of {LLM} agents,'' arXiv, 2024. [Online]. Available:
  \url{https://arxiv.org/abs/2410.09024}
\BIBentrySTDinterwordspacing

\bibitem{harmbench}
\BIBentryALTinterwordspacing
M.~Mazeika \emph{et~al.}, ``Harmbench: A standardized evaluation framework for
  automated red teaming and robust refusal,'' arXiv, 2024. [Online]. Available:
  \url{https://arxiv.org/abs/2402.04249}
\BIBentrySTDinterwordspacing

\bibitem{llmfails}
\BIBentryALTinterwordspacing
M.~Cemri \emph{et~al.}, ``Why do multi-agent {LLM} systems fail?'' arXiv, 2025.
  [Online]. Available: \url{https://arxiv.org/abs/2503.13657}
\BIBentrySTDinterwordspacing

\bibitem{nemo_guardrails}
\BIBentryALTinterwordspacing
T.~Rebedea \emph{et~al.}, ``Nemo guardrails: A toolkit for controllable and
  safe {LLM} applications with programmable rails,'' arXiv, 2023. [Online].
  Available: \url{https://arxiv.org/abs/2310.10501}
\BIBentrySTDinterwordspacing

\bibitem{mcp_spec}
\BIBentryALTinterwordspacing
{Model Context Protocol (MCP)}, ``Specification,'' Online specification, Jun.
  2025, accessed: 2025-12-17. [Online]. Available:
  \url{https://modelcontextprotocol.io/specification/2025-06-18}
\BIBentrySTDinterwordspacing

\bibitem{ncsc_prompt_injection_worse}
\BIBentryALTinterwordspacing
{UK National Cyber Security Centre (NCSC)}, ``Prompt injection is not {SQL}
  injection (it may be worse),'' Blog post, Dec. 2025, accessed: 2025-12-17.
  [Online]. Available:
  \url{https://www.ncsc.gov.uk/blog-post/prompt-injection-is-not-sql-injection}
\BIBentrySTDinterwordspacing

\bibitem{Antwi2025AdaptiveBayesianRareBinomial}
\BIBentryALTinterwordspacing
A.~Antwi, ``Adaptive bayesian interval estimation for rare binomial events,''
  \emph{Mathematics}, vol.~13, no.~24, p. 3988, 2025. [Online]. Available:
  \url{https://www.mdpi.com/2227-7390/13/24/3988}
\BIBentrySTDinterwordspacing

\bibitem{efron_tibshirani_bootstrap}
B.~Efron and R.~J. Tibshirani, \emph{An Introduction to the Bootstrap}.\hskip
  1em plus 0.5em minus 0.4em\relax Chapman and Hall/CRC, 1994.

\bibitem{chen2021_codex}
\BIBentryALTinterwordspacing
M.~Chen \emph{et~al.}, ``Evaluating large language models trained on code,''
  arXiv, 2021. [Online]. Available: \url{https://arxiv.org/abs/2107.03374}
\BIBentrySTDinterwordspacing

\bibitem{zheng2023_mtbench}
\BIBentryALTinterwordspacing
L.~Zheng \emph{et~al.}, ``Judging {LLM}-as-a-judge with {MT}-bench and chatbot
  arena,'' arXiv, 2023. [Online]. Available:
  \url{https://arxiv.org/abs/2306.05685}
\BIBentrySTDinterwordspacing

\end{thebibliography}

\end{document}